\documentclass[iop, revtex4]{emulateapj}

\usepackage{amsmath} 
\usepackage{multirow}
\usepackage{hhline}
\usepackage{xcolor}
\usepackage{rotating}
\usepackage{tablefootnote}

\newcommand\Tstrut{\rule{0pt}{2.6ex}}       
\newcommand\Bstrut{\rule[-0.9ex]{0pt}{0pt}} 
\newcommand{\TBstrut}{\Tstrut\Bstrut} 
\newcommand{\normal}{normal}
\newcommand{\bg}{91bg-like}

\shorttitle{A possible spectral sequence connecting normal and sub-luminous Type Ia Supernovae}
\shortauthors{Heringer et al.}

\begin{document}

\title{Spectral sequences of Type Ia supernovae. I. Connecting normal and sub-luminous SN Ia and the presence of unburned carbon}

\author{E. Heringer\altaffilmark{1}, M. H. van Kerkwijk\altaffilmark{1}, S. A. Sim\altaffilmark{2}, W. E. Kerzendorf\altaffilmark{3}}

\altaffiltext{1}{Department of Astronomy \& Astrophysics, University of Toronto, 50 Saint George Street, Toronto, ON, M5S 3H4, Canada}
\altaffiltext{2}{Astrophysics Research Centre, School of Mathematics and Physics, Queen’s University Belfast, Belfast BT7 1NN, UK}
\altaffiltext{3}{European Southern Observatory (ESO), Karl-Schwarzschild-Stra\ss{}e 2, D-85748 Garching, Germany}

\begin{abstract}
Type Ia supernovae are generally agreed to arise from thermonuclear explosions of carbon-oxygen white dwarfs.  The actual path to explosion, however, remains elusive, with numerous plausible parent systems and explosion mechanisms suggested. Observationally, type Ia supernovae have multiple subclasses, distinguished by their lightcurves and spectra.  This raises the question whether these reflect that multiple mechanisms occur in nature, or instead that explosions have a large but continuous range of physical properties. 
We revisit the idea that \normal\ and \bg\ supernovae can be understood as part of a spectral sequence, in which changes in temperature dominate.  Specifically, we find that a single ejecta structure is sufficient to provide reasonable fits of both the \normal\ type Ia supernova SN~2011fe and the \bg\ SN~2005bl, provided that the luminosity and thus temperature of the ejecta are adjusted appropriately. This suggests that the outer layers of the ejecta are similar, thus providing some support of a common explosion mechanism. Our spectral sequence also helps to shed light on the conditions under which carbon can be detected in pre-maximum SN~Ia spectra -- we find that emission from iron can ``fill in" the carbon trough in cool SN~Ia. This may indicate that the outer layers of the ejecta of events in which carbon is detected are relatively metal poor compared to events where carbon is not detected. 
\end{abstract}


\section{Introduction}
\label{sec:introduction}

While most type Ia supernovae (SN~Ia) are remarkably uniform, it has long been known that there are subclasses whose events share distinguishing features, such as the over-luminous ``91T-like'' and sub-luminous ``91bg-like'' \citep{Filippenko1997_spectra}.  The number of subclasses has grown with the number of available spectra, with recent reviews distinguishing almost ten \citep{Chomiuk2016_subtypes, Taubenberger2017_subtypes}.

For many of the subclasses there are associated proposed evolutionary channels and physical explosion mechanisms.  For few, however, there is consensus on which one is correct. Furthermore, just as supernova types II, IIb, Ib, and Ic have turned out to reflect not different physical origins but rather a sequence in the amount of stellar envelope stripped before core collapse, some of the observed Ia subtypes may reflect variation in explosion properties rather than different explosion mechanisms.

\citet{Nugent1995_sequence} was among the first to propose that SN~Ia belonging to different subclasses could originate from a spectral sequence in luminosity. This parameter sets the temperature profile of the ejecta, determining the ionisation state of the elements therein and thus shaping the observed spectral features. In this scenario, the amount of $^{56}$Ni synthesized in the explosion would be the primary factor driving the sequence. A similar approach was employed by \citet{Hachinger2008_tomography} to investigate in detail the strength of silicon lines.

Indeed, observationally, some subclasses seem to form continuous sequences. For instance, \citet{Branch2006_subclasses} note that their four subclasses, which are based on the pseudo equivalent widths of the $\lambda$6355 and $\lambda$5972 \ion{Si}{2} absorption features, show a continuous distribution (see Figure~\ref{Fig:parspace} below). Similarly, \citet{Benetti2005_subclasses} and \citet{Sasdelli2015_DRACULA} show that using different techniques (velocity gradient of the $\lambda$6355 \ion{Si}{2} line and principle-component analysis, respectively) the known sub-groups can be recovered but that they are not strongly disjoint, with some objects classified as ``transitional'' types.

Here, we focus specifically on the \normal\ and \bg\ subclasses. Compared to \normal\ supernovae, \bg\ events are fainter, decline faster, have a ``titanium trough'' near $4200\,\,$\AA \ and show stronger \ion{Si}{2} $\lambda$5972 and \ion{O}{1} $\lambda$7774 lines in their spectra \citep{Doull2011_91bg, Taubenberger2017_subtypes}. Moreover, the velocity inferred from the \ion{Si}{2} $\lambda$6355 line evolves more rapidly and on average settles at lower values in the post-maximum spectra \citep{Benetti2005_subclasses}. 

Also for \normal\ and \bg\ SN~Ia, it is currently unclear if they share the same progenitor scenario or arise from different channels. For instance, \citet{Sasdelli2017_subtypes} tried to classify explosion mechanisms based on their predicted spectra, and found that, while all mechanisms had shortcomings, equal-mass violent mergers fit \bg\ SN~Ia quite well, while detonations of sub-Chandrasekhar white dwarfs and delayed-detonations better represented \normal\ SN~Ia. From the distribution of inferred $^{56}$Ni masses, \citet{Piro2014_Ni} suggest that \normal\ and \bg\ SN~Ia might arise from the double detonation \citep{Livne1990_doubledet} and direct collision \citep{Dong2015_collision} scenarios, respectively. In contrast, e.g., \citet{Graur2017_subclasses} suggest a common origin on statistical grounds: while \bg\ SN~Ia are found preferentially in massive galaxies, they seem to ``cut into the share'' of the \normal\ SN~Ia.

We focus on the spectroscopic differences between two very well studied supernovae that are thought to be representative of their subclasses: the \normal\ \object{SN 2011fe} and \bg\ \object{SN 2005bl}.  For both, using the technique known as {\em abundance tomography} (e.g., \citealt{Stehle2005_tomography}), the observed spectral evolution has been used to probe different layers of the ejecta, with deeper ones exposed as the photosphere recedes.  For SN~2011fe, \citet{Mazzali2014_tomography} concluded that the progenitor likely had sub-solar metallicity, and that the density profile of the ejecta was in between the prediction from the bench-mark W7 model of \citet{Nomoto1984_W7} and the WDD1 delayed-detonation model of \citet{Iwamoto1999_WDD}.  For SN~2005bl, \citet{Hachinger2009_tomography} first optimized their synthetic spectra assuming a W7 density profile, and then used energy and mass scalings to improve their fits, finding best agreement when the original mass (of $1.38\,M_\odot$) was maintained, but the kinetic energy was reduced by 30\%.

Thus, beyond differences in explosion energy, the two analyses yielded somewhat different density profiles. Furthermore, also the abundances differ.  For instance, for titanium, the element responsible for the $4200\,$\AA \ titanium trough distinctive of \bg\ supernovae, \object{SN 2011fe} yielded a total mass $>$10 times higher than SN~2005bl in the outer layers ($v\gtrsim 7,800{\rm\,km\,s^{-1}}$), whereas SN~2005bl contains more titanium in the inner layers.

A known issue with tomography is the presence of strong degeneracies, i.e., multiple sets of parameters may give good agreement. Hence, in this first paper in a series in which we hope to investigate the observational constraints on the explosions with tomography, our primary goal is to determine whether, even though the density and abundance profiles found for both supernovae were different, the spectral differences can, in fact, be understood from variations in just one or a few parameters. We also report on a by-product of our investigation, which is that along only a limited range in our spectral sequences the carbon feature is visible; this may shed light on why carbon is only sporadically detected in SN~Ia ($\sim {\rm 30\%}$ of events with early enough spectra; e.g. \citealt{Folatelli2012_carbon}). 

\section{Methodology}
\label{sec:methodology}

To explore the physical properties that distinguish \bg\ from \normal\ SN~Ia, we use the spectral synthesis code TARDIS \citep{Kerzendorf2014_TARDIS}. TARDIS is a Monte Carlo code very similar to the code used to analyse SN~2011fe and SN~2005bl \citep{Mazzali2000_MC}, and hence our approach is to start from the available tomography \citep{Hachinger2009_tomography, Mazzali2014_tomography}. After verifying that we reproduce the earlier results, our goal is then to see for each object if we can make the synthetic spectra approach the observed spectra of the other object by a low-dimensional sequence in physical properties.

Here, for our first attempt, we follow \citet{Nugent1995_sequence} and \citet{Hachinger2008_tomography} and search for a sequence based on explosion brightness. We extend previous work by analyzing not just spectra taken near maximum light, but also before and after, and by investigating the possibility of a sequence in titanium mass fraction $X({\rm{}Ti})$.

Specifically, for our simulations we used TARDIS v1.5dev2685, which calculates a synthetic spectrum for given density and abundance profiles, an observed brightness, the time since the explosion, and the position of the pseudo photosphere (i.e., below which the ejecta are optically thick).  In TARDIS, the ejecta are assumed to be in homologous expansion, with parameters given in velocity space, and the modeled photon interactions are absorption and emission by lines (in the Sobolev approximation), and electron scattering (for details, see \citealt{Kerzendorf2014_TARDIS}).

We adopt TARDIS settings similar to those in previous works -- see appendix \ref{sec:appendix} and \citet{Kerzendorf2014_TARDIS} for a detailed explanation. For our ``default" ejecta structure of SN 2011fe we use the data available in Table 6 and Fig.~10 of \citet{Mazzali2014_tomography} and for SN 2005bl we use the same input parameters as \citet{Hachinger2009_tomography} for their best model ``05bl-w7e0.7'', as in their table A1). Starting with these, we then scale the luminosity of the explosion, following what was done by \citet{Hachinger2008_tomography} to investigate the strength of the \ion{Si}{2} $\lambda$6355 and $\lambda$5972 lines, but focusing on the spectral sequence itself, not restraining the analysis to near maximum spectra, and varying the titanium abundance as well.

For comparison purposes, all the spectra shown in this paper are normalised by the mean flux over a wavelength window of 4000--9000 $\text{\AA}$.

\section{Results}
\label{sec:results}

\begin{figure*}
\epsscale{1.1}
\plotone{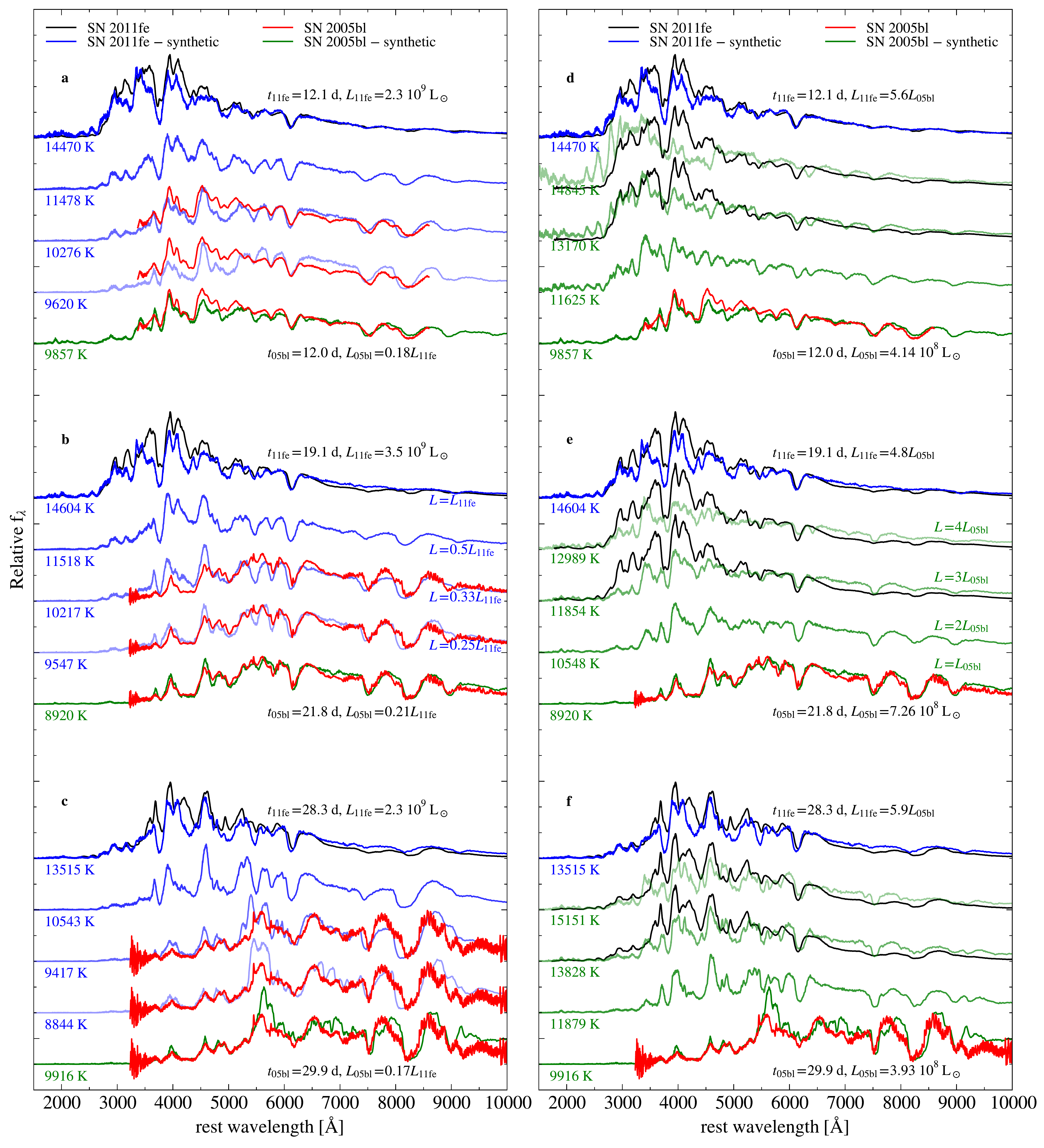}
\caption{Comparison between spectra of the \normal\ supernova 2011fe and the \bg\ supernova 2005bl with models at three different epochs. In both panels and in each set, the observed spectra for SN~2011fe and SN~2005bl are shown as black and red lines, respectively, and synthetic spectra based on literature  tomography in blue (\citealt{Mazzali2014_tomography}) and green (\citealt{Hachinger2009_tomography}), respectively. Also shown are synthetic spectra obtained with increasing changes in luminosity, in increasingly fainter lines, using the best-fit models for SN~2011fe in the \textit{left} panel, and for SN~2005bl in the \textit{right} panel.  One sees that the SN~2011fe model can reproduce the observed spectra of SN~2005bl at lower luminosity, and that, similarly, the SN~2005bl model can reproduce the observed spectra of SN~2011fe at higher luminosity.}
\label{Fig:11fe_to_05bl}
\end{figure*}

We compare our models with observed spectra\footnote{Reduced spectra were retrieved from the WISeREP archive \citep{Yaron2012_WISeREP} at \url{https://wiserep.weizmann.ac.il/}.  For details of the data reduction, see \citet{Mazzali2014_tomography} and \citet{Taubenberger2008_2005bl}.} for three epochs in Figure~\ref{Fig:11fe_to_05bl} (sets \textbf{a}--\textbf{f}). One sees that there is good agreement with the default models, showing that TARDIS allows one to reproduce the previous results, although the agreement is worse at late times, especially for SN~2005bl (see \S~\ref{sec:conclusions}).  In detail, also near and pre-maximum, our synthetic spectra are not entirely consistent with those previously published. This may reflect differences between TARDIS and the \citet{Mazzali2000_MC} code, small differences in the (interpretation of the) density and abundance profiles of SN~2011fe, and choice of atomic data. Nevertheless, the differences between the codes are no larger than those between the simulated and observed spectra, giving us confidence that relative changes are reliable.

\subsection{Luminosity scaling}

In Figure~\ref{Fig:11fe_to_05bl}, we show simulated spectral sequences in which the luminosity of the SN~2011fe and SN~2005bl models are scaled down and up, respectively, towards the luminosity of the other supernova. One sees that the observed SN~2005bl spectra are quite well reproduced by some of the fainter SN~2011fe model spectra (sets \textbf{a}--\textbf{c}.)

Indeed, our cooler SN~2011fe models not only approach the overall spectral shape of SN~2005bl, but also succeeds in reproducing some of the general spectroscopic features of \bg\ SN~Ia (see \S \ref{sec:introduction}.) For instance, the formation of the titanium trough and the stronger features due to the \ion{Si}{2} $\lambda$5972 and \ion{O}{1} $\lambda$7774 lines. This is accomplished without enhancing the abundance of these elements, but simply as an effect of the changes in ionisation states.

The above corroborates the idea that luminosity is a determining factor for the spectral differences. Looking in more detail, we see that the luminosity scale factor of $0.25$ for SN~2011fe that produces a good match to SN~2005bl near maximum corresponds to a brightness similar to the observed one for the latter. However, the same is not true for the pre- and post-maximum models (sets \textbf{a} and \textbf{c},) where the best match is for a luminosity ratio of $0.33$, while the observed ratio is about $0.18$.

In contrast, looking at the inner temperatures ($T_{\rm{}inner}$; see Fig.~\ref{Fig:11fe_to_05bl};  associated with the models, we find that near maximum, the $0.25\,L_{\rm{}11fe}$ model is closest in temperature to the default SN~2005bl model ($T_{\rm{}inner}=9547\,$K vs 8920\,K). More importantly, the temperature profile of the ejecta across the line-forming regions is similar between the scaled models for SN~2011fe and SN~2005bl. This is relevant because the precise inner temperature can vary significantly ($\pm 1000\,$K) depending on the adopted inner radius, and still produce consistent spectra. This is because the effective radius at which photons originate and the temperature at that radius are what determine the luminosity and the spectrum, and those may not be affected by the precise location of the inner boundary. This suggests the idea that rather than comparing supernovae at the same phase or brightness, one ought to compare the spectra produced when the ejecta are at a similar state, with a similar temperature profile.  To also match the luminosity, one then may have to take into account the difference in rise time ($\sim$ 17 and 19 days for SN~2005bl and SN~2011fe, respectively \citep{Taubenberger2008_2005bl, Mazzali2014_tomography}), and thus correct for the timescale on which the supernova evolves (see Sect.~\ref{sec:implications}).  

While it was previously known that the temperature is important (e.g. \citealt{Nugent1995_sequence, Mazzali1997_91bg, Branch2006_subclasses, Blondin2013_DDT},) the fact that a sequence of spectra produced from a common ejecta structure can produce both subclasses of SN~Ia by a change in temperature lends support to the idea that their outer ejecta are quite similar, despite degeneracies in, for example, the chosen density profile.

Consistent with the dominance of temperature in determining the spectral sequence is that not only do the fainter synthetic spectra based on SN~2011fe resemble SN~2005bl, the reverse is true as well: the brighter synthetic spectra based on the SN~2005bl model qualitatively resemble those of \normal\ SNe -- in particular, there is no clear 4200 \AA\ Ti trough.  In more detail, however, the cooler SN~2011fe models (sets \textbf{a}--\textbf{c}) seem to exhibit a better agreement with the observed spectra of SN~2005bl than the hotter models of SN~2005bl do when compared to SN~2011fe observations (sets \textbf{d}--\textbf{f}). This suggests that the density profile found for SN~2011fe might be closer to a common density profile for SN~Ia than the energy scaled W7 density profile employed for the SN~2005bl simulations.

Our cooler SN~2011fe models not only approach the overall spectra shape of SN~2005bl, but also succeeds in reproducing some of the general spectroscopic features of \bg\ SN~Ia (see \S \ref{sec:introduction}.) For instance, the formation of the titanium trough and the stronger features due to the \ion{Si}{2} $\lambda$5972 and \ion{O}{1} $\lambda$7774 lines. This is accomplished without enhancing the abundance of these elements, but simply as an effect of the changes in ionisation states. 

\begin{figure*}
\epsscale{1.1}
\plotone{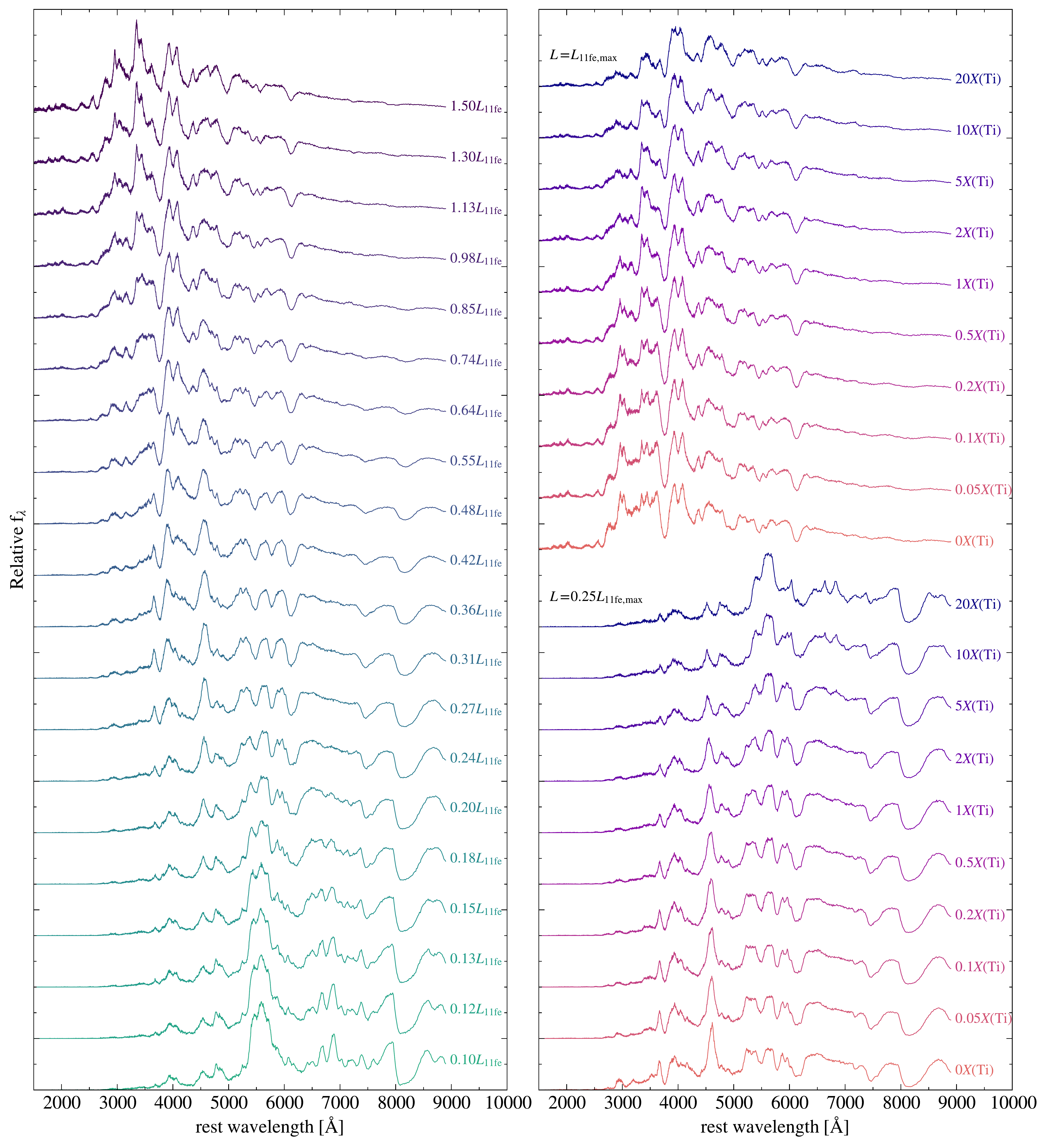}
\caption{Simulated spectra for a range in luminosity and titanium abundance.  \textit{(Left)} Varying the luminosity by a factor 0.1--1.5 in the model that best fits the maximum-light spectrum of SN~2011fe. Note how the titanium trough near $4200\,$\AA\ develops over a relatively small range in luminosity, starting to become clear at $\sim\!0.36\,L_{\mathrm{11fe}}$ and being fully formed at $\sim\!0.27\,L_{\mathrm{11fe}}$. The trough becomes less obvious at very low luminosity, where line blanketing suppresses the whole blue part of the spectrum. \textit{(Right)} Varying the titanium (and chromium; see text) abundance by a factor 0--20 for both the best model of SN 2011fe (top set) and for a model in which the luminisity is four times lower. One sees that the spectra are much more sensitive to the titanium abundance at lower luminosity.}
\label{Fig:L-grid}
\end{figure*}

\subsection{Luminosity and Titanium Sequences}

In Figure~\ref{Fig:L-grid}, we start with the default synthetic spectrum of SN~2011fe at maximum to explore how the spectra are affected in more detail by both the luminosity and thus temperature (left) and the titanium abundance (right). As a function of luminosity, one sees that the $4200\,$\AA \ titanium trough becomes progressively clearer as the luminosity decreases below  $\sim\!0.27\,L_{\mathrm{11fe}}$ and it gets blended with the \ion{Si}{2} feature near $4000\,$\AA. Other spectral features also change continuously, with, as the luminosity decreases, the \ion{Ca}{2} near-IR triplet feature near $8300\,$\AA \ becoming stronger (as fewer calcium atoms are more than singly ionized), and the \ion{Si}{2} feature near $6100\,$\AA \ first becoming stronger (similar reason) and then starting to blend with other features for $L\lesssim0.24\,L_{\mathrm{11fe}}$.

We scaled the titanium and chromium mass fractions throughout the ejecta by a factor 0--20, while imposing the condition that  $X({\rm{}Ti})=X({\rm{}Cr})$, as in \citet{Stehle2005_tomography}. Both elements are adjusted at the expense of the most abundant element in each layer (for which this is a small change: the default model contains only $\sim\!3\times 10^{-3}\,M_\odot$ of titanium above the photosphere). We use two luminosities, 1.0 and $0.25\,L_{\mathrm{11fe,max}}$, chosen such that we can both investigate how much titanium a typical \normal\ SN\ Ia could ``hide'' at maximum, without developing a Ti trough, and by what amount we can decrease titanium in cooler spectra, without loosing the trough.

From Figure~\ref{Fig:L-grid}, one sees that the amount of Ti in the ejecta will strongly influence the shape of the underlying continuum and thus the color. This effect holds for both luminosities and is as expected, since titanium and chromium, like other iron-group elements, have many transitions in the blue and ultraviolet range and thus efficiently block UV/blue light \citep{Kromer2010_doubledet}. For the spectra with $L=L_{\mathrm{11fe,max}}$ (top set), the shapes of the spectral features remain nearly unchanged, except for the \ion{Ca}{2} H and K feature near $3700\,$\AA. At $L=0.25\,L_{\mathrm{11fe,max}}$ (bottom set), the photosphere is cooler (9500\,K vs 14500\,K) and the changes in color are more drastic. The Ti trough near $4200\,$\AA \ is conspicuous as long as $X({\rm{}Ti})$ is larger than 0.05 times the default, and it disappears when Ti (and Cr) are absent from the ejecta.  For large amounts of Ti, at factors $\gtrsim\!5$ above default, the relative depth of the trough diminishes, as the spectra become redder. Overall, like \citet{Blondin2013_DDT}, we conclude that the $4200\,$\AA \ trough primarily reflects the degree of ionization of titanium.

\subsection{Carbon features}

\citet{Taubenberger2008_2005bl} report the likely detection of unburned carbon in the early spectra of SN~2005bl, due to the \ion{C}{2} $\lambda$6580 transition (seen next to the red shoulder of the silicon feature near $6100\,$\AA). While this feature is not restrained to \bg\ SN~Ia \citep{Blondin2012_review}, we note that our scaled model for SN~2011fe fails to reproduce it (see set \textbf{a}). Interestingly, this feature is present in our near maximum synthetic spectra (set \textbf{b}), but only for $L=0.50\,L_{\mathrm{11fe,max}}$. This might shed light on why carbon is only sporadically detected in the early spectra of SN~Ia. In particular, the spectral series in Fig. \ref{Fig:L-grid} reveals that the carbon signature is only clearly present in the luminosity range of $0.42\, L_{\mathrm{11fe,max}} \lesssim L \lesssim 0.98\, L_{\mathrm{11fe,max}}$.

Analysing the spectral energy distribution according to each element in our spectra, we find that the presence of the carbon trough arises from a balance between iron emission ``filling in'' the carbon trough in the cooler spectra and not enough carbon being singly ionised in the hottest cases. This finding could explain the correlation that events in which carbon is detected tend to exhibit bluer optical/near-ultraviolet colours than their counterparts in which it is not \citep{Thomas2011_carbon,Folatelli2012_carbon,Silverman2012_BSNIP_IV,Milne2012_carbon}: since iron-group elements effectively redistribute UV light to longer wavelengths, blue continua and carbon features could both be signatures of relatively low metal content (\citealt{Lentz2000_metallicity, Walker2012_metallicity}, but see \citealt{Sauer2008_metallicity}). We discuss this further in \S~\ref{sec:disc_carbon}.

\subsection{The Strengths of the \ion{Si}{2} $\lambda$6355 and $\lambda$5972 features}

To put the effects of changes in luminosity and temperature in a larger context, we investigated how they affect features that have been used to classify SN~Ia.  Specifically, we use the strong features near 6100 and $5700\,$\AA\ -- due to \ion{Si}{2} $\lambda$6355 and $\lambda$5972 -- which define the \textit{core-normal}, \textit{shallow-silicon}, \textit{cool} and \textit{broad-line} subclasses of \citet{Branch2006_subclasses, Branch2009_subclasses}. Here, the \textit{cool} subclass overlaps with the \textit{faint} group defined by \citet{Benetti2005_subclasses}, which we generically call \bg. For each spectrum shown in the left panel of Fig. \ref{Fig:L-grid}, we compute pseudo-equivalent widths (pEW) following the prescription of \citet{Silverman2012_BSNIP_II}. We estimate uncertainties following \citet{Liu2016_pEW}, using a Monte Carlo routine to generate mock spectra with noise properties like the data (estimated from by smoothing the spectrum and computing the root mean square between the smoothed spectrum and the data in a wavelength window of $40\,$\AA).  By analysing multiple TARDIS runs with identical inputs, we found the resulting uncertainties slightly underestimate the true ones by $\sim\!20$\% (for which we correct).

\begin{figure}
\epsscale{1.1}
\plotone{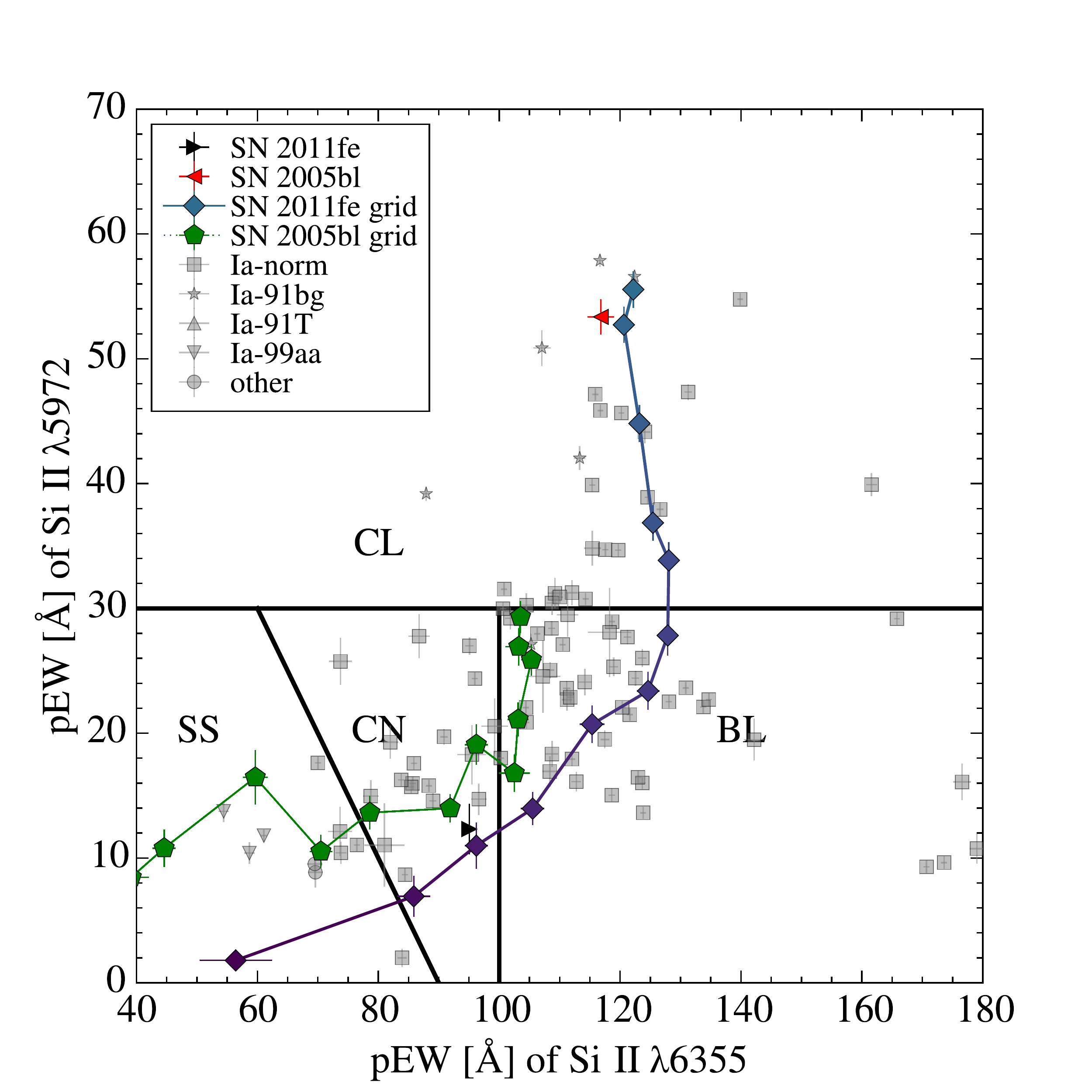}
\caption{Observed and simulated pseudo-equivalent widths of the \ion{Si}{2} $\lambda$6355 and $\lambda$5972 features (after Fig.~10 of \citealt{Silverman2012_BSNIP_II}). The four regions indicate the four subclasses identified by \citet{Branch2009_subclasses}: core-normal (CN), shallow-silicon (SS), cool (CL) and broad-line (BL).  Observed widths (gray markers) are for objects from the BSNIP survey \citep{Silverman2012_BSNIP_I}, and simulated widths are for spectra based on the best-fit model for SN~2011fe at maximum, but with the luminosity scaled by a factor 0.31--1.5 (connected diamonds, color-coded as in the left panel of Fig.~\ref{Fig:L-grid}), and for SN~2005bl, scaled by a factor 1.13--7.2 (green pentagons); for even lower luminosities, features blend, making it difficult to measure the pseudo-equivalent widths).}
\label{Fig:parspace}
\end{figure}

We remeasured pEWs of objects from the Berkeley Supernova Ia Program (BSNIP; \citealt{Silverman2012_BSNIP_I}) and show them in Figure~\ref{Fig:parspace} (which can be compared with Fig. 10 from \citealt{Silverman2012_BSNIP_II}; our pEW are consistent).  Overlaid are the pEW we measured from the synthetic spectra based on the SN~2011fe model at maximum, but with scaled luminosity.  One sees that the spectral sequence crosses through all the four subclasses displayed.  The brightest explosions are in the \textit{shallow-silicon} region, reflecting that most of the silicon is more than singly ionized at the corresponding high temperatures.  Towards lower brightness, the fraction of singly-ionised Si increases strongly (while only a negligible fraction is neutral) and the \ion{Si}{2} $\lambda$6355 feature strengthens until it saturates near a value of $\sim\!130\,$\AA, while the \ion{Si}{2} $\lambda$5972 feature becomes progressively stronger \citep{Hachinger2008_tomography}.

\section{Modelling uncertainties}

An important limitation of our analysis follows from the approximation of the incoming radiation field from the pseudo-photosphere as black-body radiation. This approximation becomes progressively worse starting about a week after maximum, as the ejecta evolve and the photosphere moves in too far. For comparison with previous work, we nevertheless included some simulations at such late epochs.

It is generally suggested that the \texttt{macroatom} TARDIS mode for line treatment is used instead of the \texttt{downbranch} mode that we employ, since the former provides a more physical approximation. 
Specifically, the \texttt{macroatom} approximation better describes deexcitation cascades (and multiple-photon excitation, though those are less likely to be important) than does the simpler \texttt{downbranch} mode.
Nevertheless we retain the \texttt{downbranch} approximation for consistency with the earlier abundance tomography models on which our study is based: to consistently use the \texttt{macroatom} mode would require re-doing the tomography of SN~2011fe and SN~2005bl, which is beyond the scope of this work.

\begin{figure}
\epsscale{1.1}
\plotone{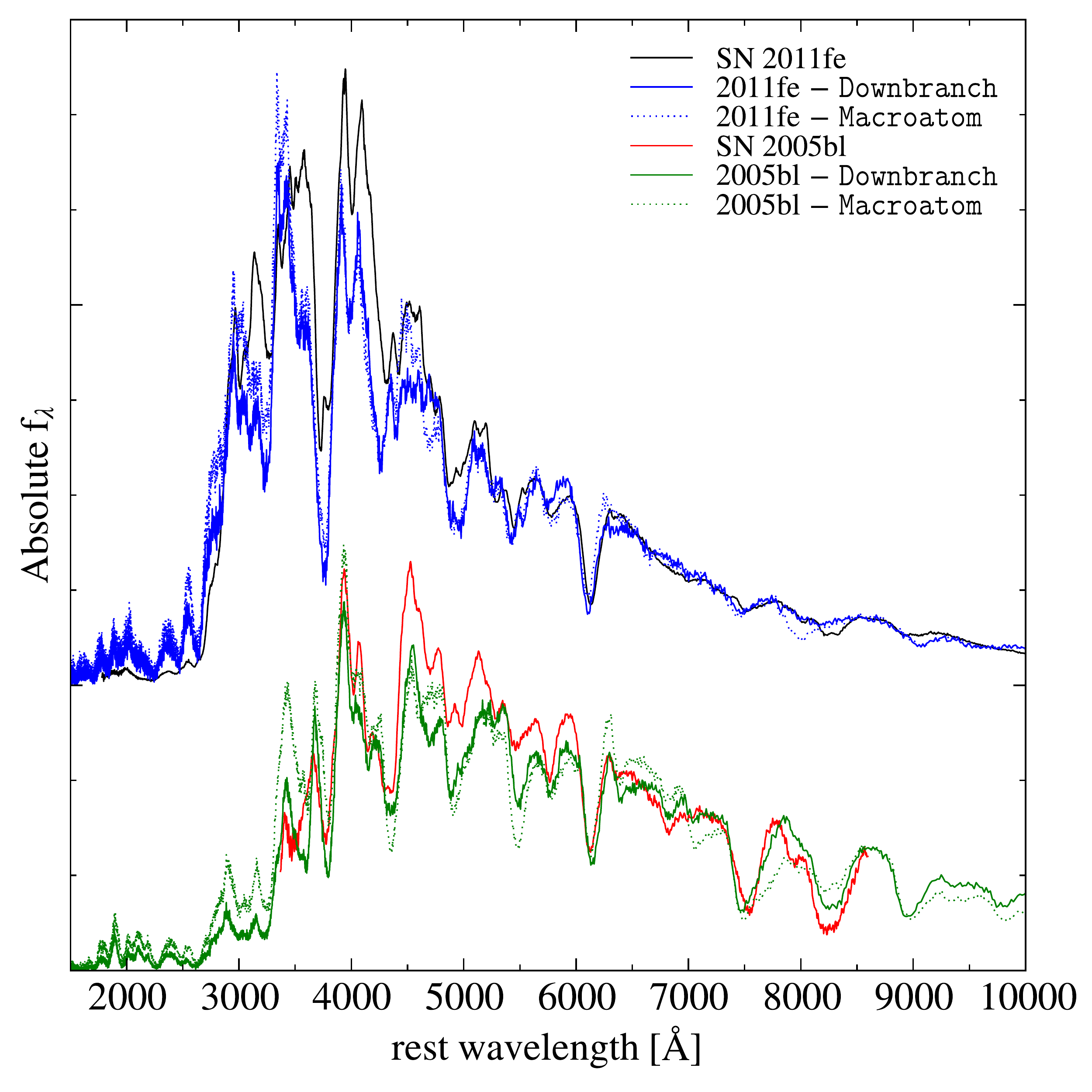}
\caption{Comparison between synthetic spectra computed using the \texttt{downbranch} (full line) and \texttt{macroatom} (dashed line) line interaction modes. For reference, the observed spectra of SN~2011fe (black) and of SN~2005bl (green) are also shown, as in Fig. \ref{Fig:11fe_to_05bl}, set \textbf{a}.}
\label{Fig:line_comparison}
\end{figure}

However, while \cite{Kerzendorf2014_TARDIS} found few substantial differences between \texttt{macroatom} and \texttt{downbranch} calculations for the model they considered, our particular model is noticeably affected by this choice. In Fig. \ref{Fig:line_comparison}, we show synthetic spectra for the ``default'' model for the early spectra of SN~2011fe and SN~2005bl (set \textbf{a} in Fig.~\ref{Fig:11fe_to_05bl}), computed using both the \texttt{macroatom} and \texttt{downbranch} modes. Looking at the simulations in detail, we find that the emission and absorption of Si and Mg in particular differs substantially between the two treatments, with cooler spectra more strongly affected. Clearly, this is important for detailed fits, and we hope to investigate the origin of this difference in the future.

To help give a sense of the differences, we have added an extra version of Figs. \ref{Fig:11fe_to_05bl}--\ref{Fig:parspace} in the appendix \ref{sec:appendix}, in which the synthetic spectra are computed with the \texttt{macroatom} line interaction mode. While our main conclusions are not affected by these differences, the comparison between SN~2005bl and a ``fainter'' version of SN~2011fe are less strikingly similar. Of course, for these simulations, the starting models are not optimal, since the tomography was done using a different treatment. It will require tomography using the \texttt{macroatom} mode, however, to test whether starting from proper initial conditions, the spectra would look more similar again.

\section{Conclusions}
\label{sec:conclusions}
\subsection{Implications for the \normal\ and \bg\ subtypes}
\label{sec:implications}

We have shown that for the same ejecta structure, synthetic spectra can reproduce observations of both the \normal\ SN~Ia SN~2011fe and the \bg\ SN~2005bl, with only the input luminosity adjusted such that the temperatures in the ejecta are similar to those found in earlier fits in the literature. In particular, the titanium trough and the evolution of the strength of the silicon, oxygen and calcium features are in agreement with the data. The luminosities themselves are not always consistent with the observed ones. One would likely obtain better agreement with the observed spectra at the correct luminosity if one were to take into account that SN~2005bl evolved substantially faster than SN~2011fe ($\Delta m_{\mathrm{15, 05bl}} = 1.93 \pm 0.10$ mag \citep{Taubenberger2008_2005bl} vs. $\Delta m_{\mathrm{15, 11fe}} = 1.07 \pm 0.06$ mag \citep{McClelland2013_11fe}). From initial experiments, we find that there is a degeneracy between the time since explosion and the luminosity, i.e., reducing the value of time since explosion in the calculations for the scaled down SN~2011fe model, we can approach the temperature structure and thus the SN~2005bl spectrum similarly well for the correct luminosity.

We have also investigated the effects of changing the titanium and chromium mass fractions, finding that the spectra are relatively insensitive to these: starting with the SN~2011fe model at maximum, abundance changes by a factor 0.05--20 influence the color but do not affect the shape of the Ti trough much. At lower luminosity, a Ti trough is a generic feature, and will exist even for 20 times lower than the nominal abundance of $\sim\!3\times 10^{-3}\,M_\odot$ above the photosphere \citep{Mazzali2014_tomography}. Overall, we conclude that \normal\ SN~Ia do not have to be Ti poor compared to their \bg\ counterparts, but simply hotter, causing titanium and chromium to be doubly ionized, thus preventing the formation of the Ti trough (as also found by \citealt{Blondin2013_DDT}.)

Our results seem suggestive of a common explosion mechanism for \bg\ and \normal\ SN~Ia, with the subclasses arising from a continuous set of parameters within a single mechanism.  Of course, it does not directly distinguish between mechanisms: indeed, both have been reproduced by delayed detonation \citep{Blondin2013_DDT}, violent mergers \citep{Pakmor2010_DD, Pakmor2012_DD} and double detonation \citep{Shen2017_doubledet} models.

A remaining caveat is that, since the spectra of SN~2005bl can be reproduced by models with distinct ejecta profiles, we cannot truly determine whether its ejecta structure was similar to that of SN~2011fe; in the end, it is mostly an appeal to Occam's razor.  Similarly, our results do not exclude potential subdivisions within each of the subclasses.  For instance, \citet{Dhawan2017_subclasses} used the epoch of maximum $B-V$, whether one or two peaks are seen in the near-infrared, and the epoch of the first near-infrared peak relative to that in the blue, to argue that there are two distinct groups of \bg\ SN~Ia, of which one exhibits a continuous set of properties with \normal\ SN~Ia, while the other does not.  SN~2005bl was part of their sample, and was classified as belonging to the second, unconnected group, although lack of some of the photometric evidence made the classification tentative.

Fortunately, the similarity can be tested: e.g., if indeed \bg\ SN~Ia have ejecta structure similar to that of \normal\ SN~Ia, it is possible to predict how their early spectra -- which have not yet been observed -- will look.  Furthermore, with updated simulations, it should be possible to extend the comparison to later times (e.g. \citealt{Mazzali2015_2011Fe, Botyanszki2017_nebular}), when the inner parts of the ejecta become visible, which surely should be different: \bg\ SN~Ia must necessarily produce less $^{56}$Ni.

More generally, our investigation could be extended to other SN~Ia subtypes, such as the overluminous 91T-like objects, as well as to objects classified as ``transitional'' between \normal\ and \bg\ SN~Ia, such as SN~1986G \citep{Ashall2016_tomography} and SN~2004eo \citep{Pastorello2007_2004eo, Mazzali2008_tomography}.  At the same time, Fig.~\ref{Fig:parspace} shows that observed SN~Ia do not follow a one-parameter family; it will be interesting to explore what other properties need to be changed to reproduce the SN~Ia more distant from our models. 

\subsection{Metal content in the outer layers}
\label{sec:disc_carbon}

Our cooler models indicate that the detection of the carbon feature near $6100\,$\AA\ might be hindered by emission from iron, even at pre-maximum, where it had not hitherto been considered (cf., \citealt{Branch2008_postmaximum, Parrent2012_carbon}). This suggests that the carbon trough might preferentially be seen in low metallicity progenitors or in strongly stratified objects, where no or very little of the iron-group elements formed deeper in the ejecta are mixed up to the outer layers.

It is worth noting that \bg\ SN~Ia occur preferentially in massive  elliptical galaxies (e.g. \citealt{Taubenberger2017_subtypes}), which tend to be relatively metal rich. Therefore, it might be the case that the same mechanism that causes such explosions to produce less $^{56}$Ni, also affects the formation of the carbon trough.

Carbon was also detected in SN~2011fe \citep{Nugent2011_WD,Parrent2012_carbon}. For SN~2011fe, a sub-solar metallicity was inferred from the tomography of \citet{Mazzali2014_tomography}, as well as from comparisons with its ``twin,'' SN~2011by (which also exhibits a carbon trough), with  \citet{Folley2013_metallicity} concluding SN~2011fe had a lower luminosity, and \citet{Graham2015_metallicity} noting that the distance to SN~2011by might be underestimated, but that then both these SNe~Ia would arise from low metallicity progenitors. In contrast to this, however, \citet{Baron2015_11fe} found that, in the context of delayed detonations models, the strength of the carbon feature is more easily reproduced using solar than sub-solar abundances.

Yet another example that might help to elucidate the relationship between the carbon trough and the Fe content of the ejecta is SN~iPTF13asv. This overluminuous event exhibited a weak (but persistent) carbon signature, while the early spectra showed an absence of Fe features. In addition, this explosion was UV bright near maximum, its ejecta was clearly stratified and it originated from a metal-poor host galaxy \citep{Cao2016_subtype}.

Fortunately, this also is amenable to verification.  With more detailed models, it should be possible to understand exactly where in the ejecta the presence of iron-group elements makes the carbon feature harder to detect. Furthermore, observationally, it should be possible to determine whether the carbon feature is easier to detect for SNe~Ia from low-metallicity progenitors by correlating with the metallicity of environments as measured by, e.g., \citet{Anderson2015_young}.

\bigskip
\bigskip

We thank Melissa L. Graham for meaningful discussions about the carbon feature in SN~Ia and its possible connection with the metallicity of the progenitor star and Stephan Hachinger for providing us with the input parameters used in their tomography work of SN~2005bl.

\appendix

\section{Appendix A: Simulation parameters}
\label{sec:appendix}

Here we present the set of parameters used in our simulations. We attempted to reproduce the tomography analysis of SN~2011fe, as published by \citet{Mazzali2014_tomography}, and of SN~2005bl, published by \citet{Hachinger2009_tomography}. Specifically, to make our assumptions as similar as possible to those used in previous work, we adopt the \texttt{downbranch}, \texttt{dilute-blackbody}, \texttt{dilute-lte} and \texttt{nebular} modes to treat line interaction, radiative rates, excitation and ionization, respectively. 
The model inputs and the temperature at the inner boundary are given in Table \ref{tb:simulation}, while the abundance stratification adopted is given in Table \ref{tb:abundance}. For comparison purposes, we have also run the simulations adopting the more physical line interaction mode, \texttt{macroatom}, while keeping all other parameters unchanged. While the differences in the computed spectra are not negligible, they do not affect our main conclusions.

\onecolumngrid
\newpage

\begin{deluxetable}{cccccc}
\tablecaption{Simulation properties. \label{tb:simulation}}
\tablehead{
\multirow{2}{*}{Model} & t & lg $L$ & $v_{\mathrm{inner}}$ & $T_{\mathrm{inner}}$\tablenotemark{a} & $T_{\mathrm{inner}}$\tablenotemark{b} \TBstrut \\
 & (d) & ($L_{\mathrm{bol}}/L_\odot$) & (km s$^{\mathrm{-1}}$) & (K) & (K)\TBstrut \Bstrut
}
\startdata
\multirow{5}{*}{05bl} & -6 & 8.520 & 8350 & 9765 & 9639 \\ 
 & -5 & 8.617 & 8100 & 10\,136 & 9857 \\ 
 & -3 & 8.745 & 7600 & 10\,633 & 10463 \\
 & 4.8 & 8.861 & 6800 & 8931 & 8920 \\
 & 12.9 & 8.594 & 3350 & 10072 & 9916\Bstrut  \\
 \hline
\multirow{8}{*}{11fe} & 3.7 & 7.903 & 13\,300 & 10\,800 & 10\,789\Tstrut \\
 & 5.9 & 8.505 & 12\,400 & 12\,100 & 12\,012 \\
 & 9.0 & 9.041 & 11\,300 & 14\,500 & 13\,974 \\
 & 12.1 & 9.362 & 10\,700 & 14\,900 & 14\,470 \\
 & 16.1 & 9.505 & 9000 & 15\,100 & 14\,856 \\
 & 19.1 & 9.544 & 7850 & 14\,700 & 14\,604 \\
 & 22.4 & 9.505 & 6700 & 14\,100 & 13\,901 \\
 & 28.3 & 9.362 & 4550 & 13\,500 & 13\,515
\enddata
\tablenotetext{a}{Temperature at the inner radius as given in the literature (\citealt{Hachinger2009_tomography,Mazzali2014_tomography}.)}
\tablenotetext{b}{Temperature at the inner radius obtained from the simulations used in this work.}
\end{deluxetable}

\begin{sidewaystable}[ph]
\begin{center}
\caption{Abundance stratification. \label{tb:abundance}}
\begin{tabular}{cccccccccccccc}
\hline\hline
Model & $v$ (km s$^{\mathrm{-1}}$)  & $X$(C) & $X$(O) & $X$(Na) & $X$(Mg) & $X$(Al) & $X$(Si) & $X$(S) & $X$(Ca) & $X$(Ti) & $X$(Cr) & $X$(Fe)$_\mathrm{0}$\tablenotemark{a} & $X$(Ni)$_\mathrm{0}$\tablenotemark{a} \TBstrut \\
\hline\hline
\multirow{5}{*}{05bl} & 16\,000 $-$ 33\,000 & 0.4184 & 0.5726 & 0.0000 & 0.0030 & 0.0000 & 0.0050 & 0.0010 & 0.0000 & 0.0000 & 0.0000 & 0.0000 & 0.0000\Tstrut \\ 
 & 8400 $-$ 16\,000 & 0.0600 & 0.8600 & 0.0060 & 0.0400 & 0.0025 & 0.0200 & 0.0100 & 0.0004 & 0.0004 & 0.0003 & 0.0003 & 0.0000\Tstrut \\ 
 &  8100 $-$ 8400 & 0.0300 & 0.1300 & 0.0030 & 0.0300 & 0.0025 & 0.6800 & 0.1000 & 0.0004 & 0.0100 & 0.0070 & 0.0150 & 0.0000 \\ 
 & 7500 $-$ 8100 & 0.0000 & 0.0100 & 0.0000 & 0.0000 & 0.0015 & 0.7000 & 0.1100 & 0.0004 & 0.0533 & 0.0367 & 0.0900 & 0.0000 \\
 & 6600 $-$ 7500 & 0.0000 & 0.0000 & 0.0000 & 0.0000 & 0.0000 & 0.7100 & 0.0700 & 0.0004 & 0.0550 & 0.0400 & 0.1150 & 0.0100 \\
 & 3300 $-$ 6600 & 0.0000 & 0.0000 & 0.0000 & 0.0000 & 0.0000 & 0.7700 & 0.0000 & 0.0005 & 0.0167 & 0.0167 & 0.0650 & 0.1300\Bstrut  \\
\hline
\multirow{11}{*}{11fe} & 19500 $-$ 24000  & 0.9804  & 0.0120  & 0.0000  & 0.0030  & 0.0000  & 0.0040  & 0.0005  & 0.0000  & 0.0000  & 0.0000  & 0.0001  & 0.0000  \\
 & 16000 $-$ 19500  & 0.0260  & 0.8605  & 0.0000  & 0.0300  & 0.0000  & 0.0600  & 0.0200  & 0.0020  & 0.0000  & 0.0000  & 0.0005  & 0.0010  \\
 & 13500 $-$ 16000  & 0.0310  & 0.7030  & 0.0000  & 0.0300  & 0.0000  & 0.2000  & 0.0300  & 0.0030  & 0.0005  & 0.0005  & 0.0010  & 0.0010  \\
 & 12000 $-$ 13500  & 0.000  & 0.351  & 0.000  & 0.020  & 0.000  & 0.440  & 0.080  & 0.003  & 0.003  & 0.003  & 0.060  & 0.040  \\
 & 11000 $-$ 12000  & 0.008  & 0.110  & 0.000  & 0.000  & 0.000  & 0.563  & 0.150  & 0.003  & 0.005  & 0.005  & 0.006  & 0.150  \\
 & 9000 $-$ 11000  & 0.0080  & 0.0900  & 0.0000  & 0.0000  & 0.0000  & 0.4785  & 0.1500  & 0.0030  & 0.0050  & 0.0050  & 0.0005  & 0.2600  \\
 & 8500 $-$ 9000  & 0.0080  & 0.0900  & 0.0000  & 0.0000  & 0.0000  & 0.1985  & 0.0700  & 0.0030  & 0.0050  & 0.0050  & 0.0005  & 0.6200  \\
 & 8000 $-$ 8500  & 0.0080  & 0.0200  & 0.0000  & 0.0000  & 0.0000  & 0.2285  & 0.0700  & 0.0030  & 0.0050  & 0.0050  & 0.0005  & 0.6600  \\
 & 7500 $-$ 8000  & 0.0000  & 0.0000  & 0.0000  & 0.0000  & 0.0000  & 0.2165  & 0.0700  & 0.0030  & 0.0050  & 0.0050  & 0.0005  & 0.7000  \\
 & 7000 $-$ 7500  & 0.0000  & 0.0000  & 0.0000  & 0.0000  & 0.0000  & 0.2000  & 0.0600  & 0.0030  & 0.0050  & 0.0050  & 0.0005  & 0.7265  \\
 & 3500 $-$ 7000  & 0.0000  & 0.0000  & 0.0000  & 0.0000  & 0.0000  & 0.0900  & 0.0300  & 0.0001  & 0.0010  & 0.0010  & 0.1500  & 0.7279  \\\Bstrut \\
\hline
\tablenotetext{a}{Fe and Ni mass fractions are, for convenience, given at $t\,=\,0$, before the decay rates are taken into account.}
\end{tabular}
\end{center}
\end{sidewaystable}

\newpage

\begin{figure*}
\epsscale{1.1}
\plotone{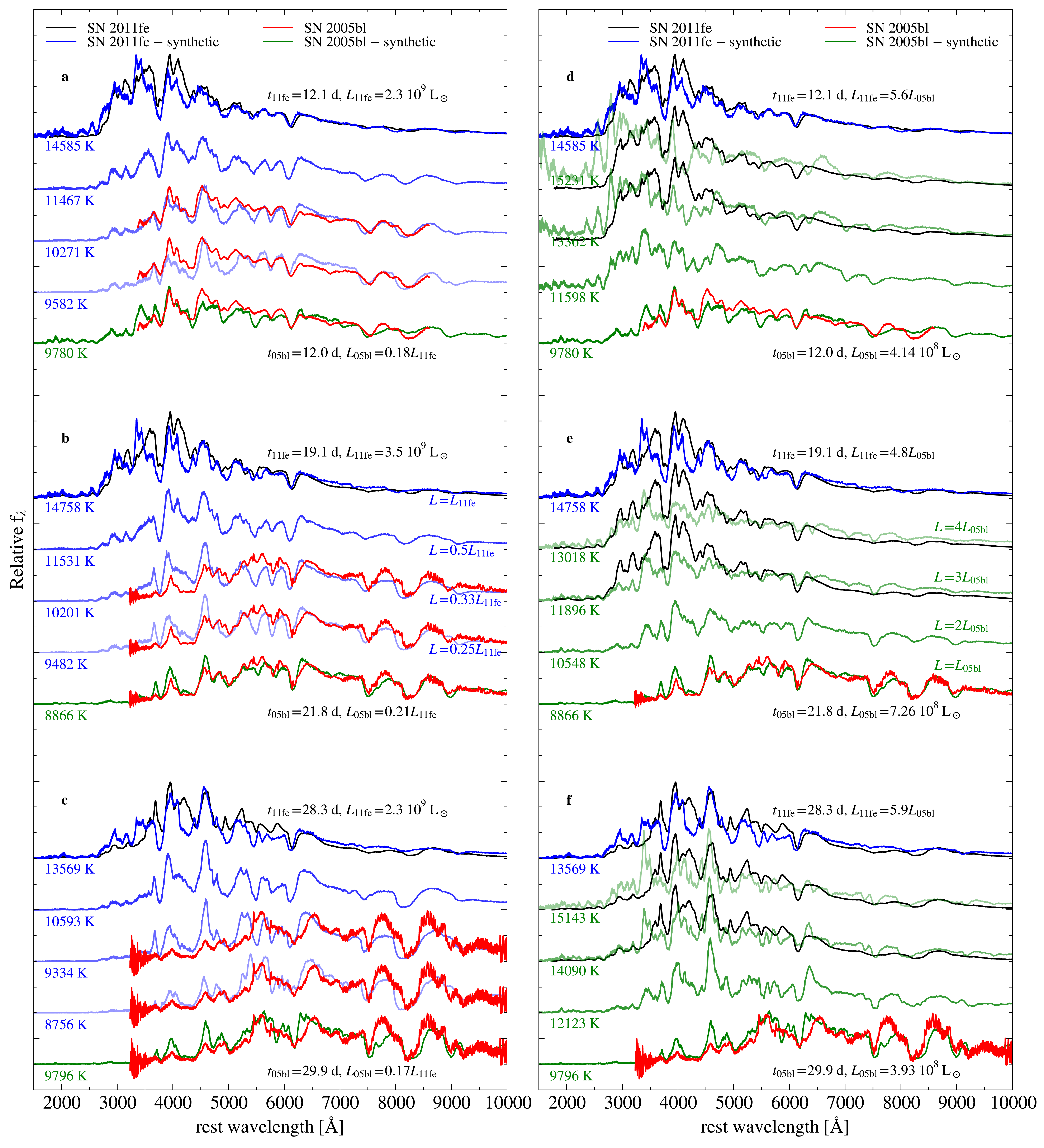}
\caption{Same as Fig. \ref{Fig:11fe_to_05bl}, but the simulations adopt the \texttt{macroatom} line interaction mode.}
\label{Fig:11fe_to_05bl_macroatom}
\end{figure*}

\begin{figure*}
\epsscale{1.1}
\plotone{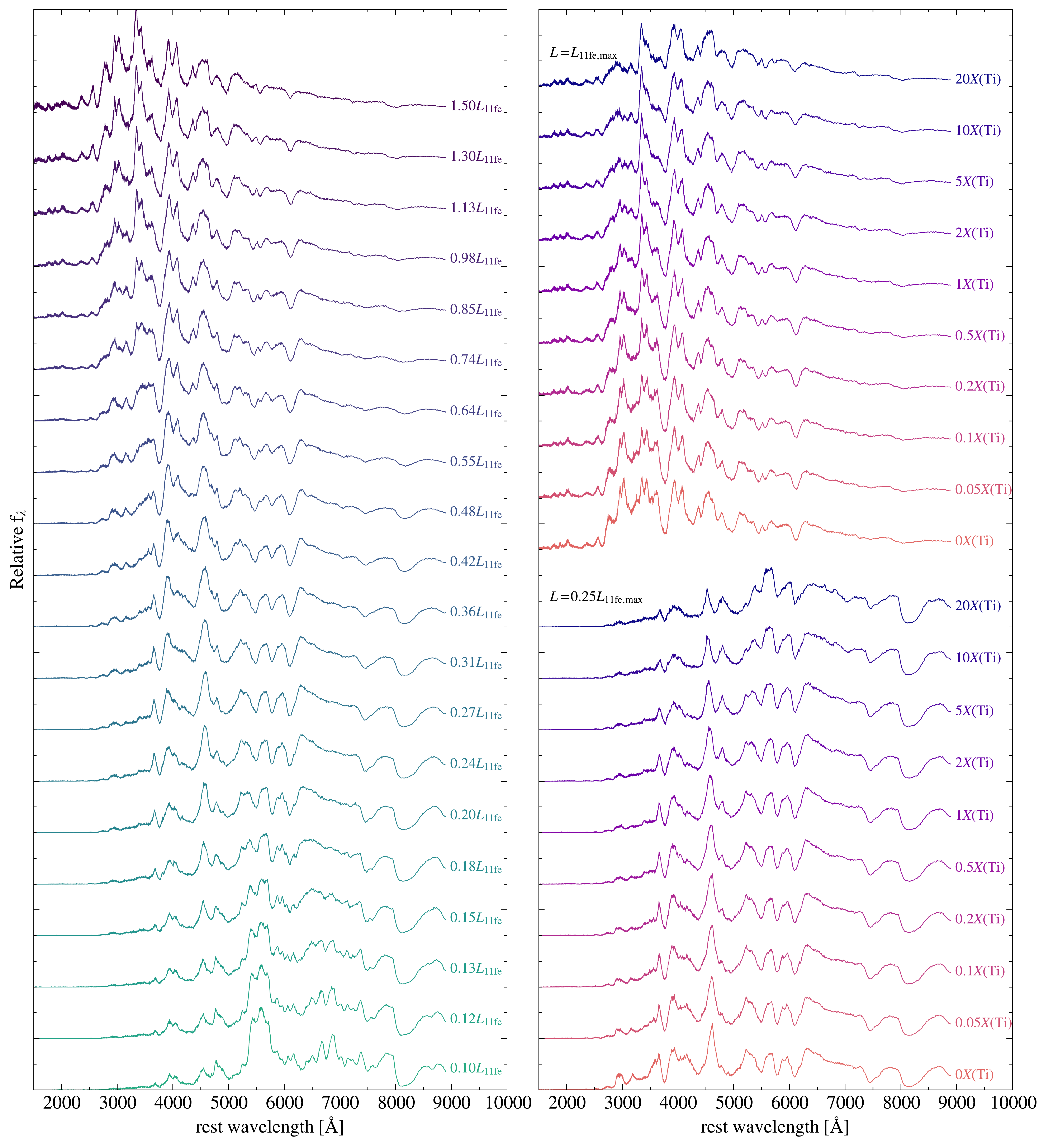}
\caption{Same as Fig. \ref{Fig:L-grid}, but the simulations adopt the \texttt{macroatom} line interaction mode.}
\label{Fig:L-grid_macroatom}
\end{figure*}

\begin{figure}
\epsscale{1.1}
\plotone{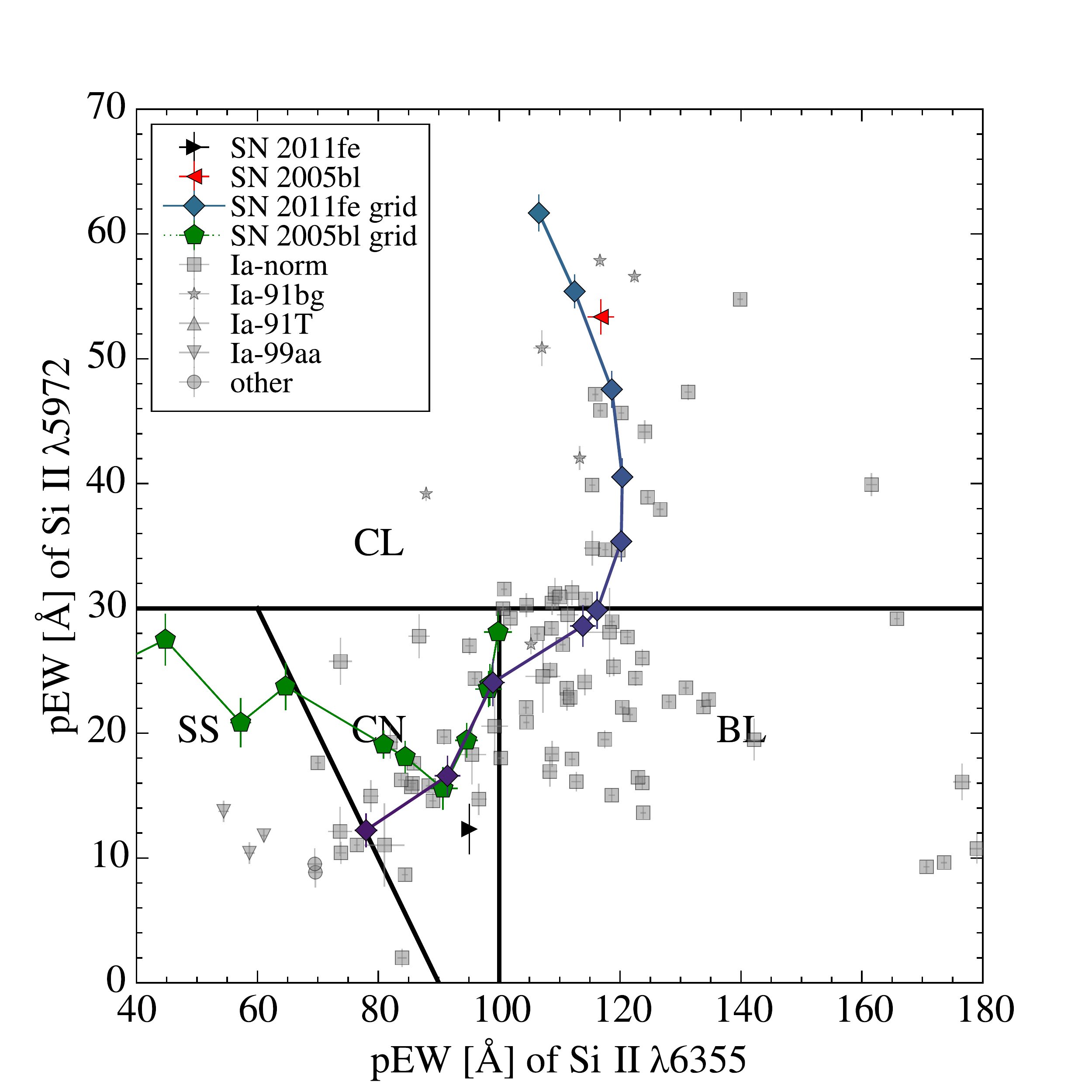}
\caption{Same as Fig. \ref{Fig:parspace}, but the simulations adopt the \texttt{macroatom} line interaction mode.}
\label{Fig:parspace_macroatom}
\end{figure}

\end{document}